\documentstyle[times,amssymb,amsfonts,latexsym,preprint,eqsecnum,aps,pra,12pt]{revtex}
\tighten
\draft
\def\vec#1{\bbox{\rm #1}}
\begin{document}

\title{\bf BORROMEAN BINDING OF THREE OR FOUR  BOSONS}
\author{Steven Moszkowski}         
 \address{
 Department of Physics and Astronomy, UCLA\\
 Los Angeles, CA 90095, USA}
\author{Sonia Fleck  and Ali Krikeb} 
\address{
Institut de Physique Nucl\'eaire de Lyon, Universit\'e 
Claude Bernard,
CNRS-IN2P3\\
43, Boulevard du 11 Novembre 1918, 
F-69622 Villeurbanne Cedex, France}
\author{Lukas Theu{\ss}l and Jean-Marc Richard}
\address{ 
Institut des Sciences Nucl\'eaires, Universit\'e Joseph Fourier, CNRS-IN2P3\\
53, Avenue des Martyrs, F-38026 Grenoble Cedex, France}
\author{K{\'a}lm{\'a}n Varga} 
\address{Argonne National Laboratory\\
9700 S.Cass Ave, Argonne, IL 60439, USA}
\preprint{
\begin{tabular}[h]{r}ISN--00-18\\LYCEN 2000/10
\end{tabular}}
\maketitle 
\begin{abstract}
We estimate the ratio $R=g_{3}/g_{2}$ of the critical coupling 
constants $g_{2}$ and $g_{3}$ which are required to achieve binding of 
2 or 3 bosons, respectively, with a short-range interaction, and 
examine how this ratio depends on the shape of the potential.  Simple 
monotonous potentials give $R\simeq 0.8$.  A wide repulsive core 
pushes this ratio close to $R=1$.  On the other hand, for an 
attractive well protected by an external repulsive barrier, the ratio 
approaches the rigorous lower bound $R=2/3$.  We also present results 
for $N=4$ bosons, sketch the extension to $N>4$, and discuss various 
consequences.
\end{abstract}
\section{Introduction}\label{Intro}
The phenomenon of ``Borromean'' binding is well-known 
\cite{Bang93,Riisager94}.  In our world with 3 dimensions, a 
short-range potential $g v(r)$ does not always achieve binding of two 
bodies, even if $v$ is attractive or contains attractive parts.  A 
minimal strength $g_{2}$ is needed, more precisely $m g\ge g_{2}$, 
where $m$ is the mass of the constituents.  Similarly, binding three 
identical bosons of mass $m$ requires $mg\ge g_{3}$ for the pairwise 
interaction $g\sum v(r_{ij})$, where $r_{ij}$ denotes the distance 
between particles $i$ and $j$.  The crucial observation is that 
$g_{3}< g_{2}$, implying that for a coupling $g$ such that 
$g_{3}<g<g_{2}$, the 3-body system is bound while none of its 
subsystems is bound.

An example is the $^{6}\mathrm{He}$ nucleus, considered schematically 
as a $(\alpha,\,n,\, n)$ system.  It is stable against spontaneous 
dissociation, while neither the $(\alpha,\,n)= {}^{5}\mathrm{He}$ nor 
the $(n,\,n)$ systems are bound.  The name ``Borromean'' was given to 
such nuclei after the Borromean rings, which are interlaced is such a 
subtle topological way, that when one removes one of them, the two 
others become unlocked.

Borromean binding is implicit to understand the Thomas collapse 
\cite{Thomas35}.  When the range of the potential $v$ is reduced, or 
equivalently, when $g\to g_{2}$ from above, the 3-body binding energy 
$E_{3}(g)$ becomes very large compared to the 2-body energy $E_{2}$.  
Also the Efimov effect \cite{Efimov70}, i.e., the proliferation of 
loosely-bound excited states in the 3-body spectrum near $g=g_{2}$ 
implies that the 3-body ground-state already exists at this point.

Let $R=g_{3}/g_{2}$ be the ratio of critical coupling constants.  For simple 
monotonous potentials, such as Yukawa, Gaussian or exponential, it is 
found \cite{Richard94} that $R$ is very close to $0.8$.  This is in 
agreement with the rigorous lower bound $R\ge 2/3$ \cite{Richard94}.  
The fact that all simple potentials give almost the same $R\simeq 0.8$ 
is understood as follows: at vanishing energy, the wave function 
extends very far outside the potential well, and thus does not probe 
very accurately the details of the short-range interaction, which is 
just seen as a contact attraction.

 There are, however, reasons to believe that $R$ can appreciably 
 differ from 0.8.  The aim of the present paper is precisely to study 
 how $R$ evolves when one starts from a simple monotonous potential and 
 adds either an inner core or an external barrier.

 When an external barrier of growing size is added to the potential, 
 $R$ evolves from $R\simeq 0.8$ to $R\to 2/3$.  An example is provided 
 by $v\propto r^{2}\exp(-2 \mu r)-\exp(-\mu r)$ when $\mu$ varies, or, 
 similarly, by combinations of Gaussians.

 When an inner core is implemented, a transition is observed from 
 $R\simeq 0.8$ to $R\to 1$.  This will be seen for the Morse and the 
 P{\"o}schl--Teller potentials when an appropriate parameter is 
 varied.  An extreme case consists of an hard core of radius $c$ and 
 an attractive delta-shell $-\delta(r-d)$ located at $d>c$.  The 
 critical strength $g_{2}$ can be calculated exactly.  One can also 
 calculate exactly the strength $g_{\infty}$ which makes the 2-body 
 scattering length vanish and hence is sufficient to bind the infinite 
 boson matter \cite{Bruch76}, with the result $g_{\infty}/g_{2}=c/d$.  
 Thus, as $d\to c$, $g_{\infty}/g_{2}$ approaches 1 and so does any 
 $g_{N}/g_{2}$ ratio with $N$ finite.

 Note that the ratio $R$ cannot exceed $R=1$ for the additive 
 potential V=$g\sum v(r_{ij})$, provided $v$ is purely attractive.  
 This means that one cannot conceive a situation where 2-body systems 
 are bound while a 3-boson system is unbound.  The following proof is 
 due to Basdevant \cite{Basdevant98}.  For $g<g_{2}$, let $\varphi(r)$ 
 be the ground state wave-function of the 2-body system, with energy 
 $E_{2}$.  The trial wave-function $\Psi= 
 \varphi(r_{12})\varphi(r_{13})$ can be used for the 3-body Hamiltonian 
 written as,
\begin{equation}\label{eq:3-body-H}
	H={\vec{p}_{2}^{2}\over 2}\left({1\over m_{1}}+{1\over m_{2}}\right)
	+{\vec{p}_{3}^{2}\over 2}\left({1\over m_{1}}+{1\over m_{3}}\right)
	+ {\vec{p}_{2}\!\cdot\!\vec{p}_{3}\over 
	m_{1}}+v(r_{12})+v(r_{23)}+v(r_{31}),
\end{equation}
 leading to an expectation value $2 E_{2}$ if the interaction $v(r_{23})$ 
 is neglected.  So $E_{3}\le 2E_{2}<0$ if $v\le 0$, q.e.d.\@ The 
 proof holds for an asymmetric interaction V=$\sum v_{ij}$ with 
 $v_{12}$ binding $(1,\,2)$ and $v_{13}$ binding $(1,\,3)$, and 
 $v_{23}$ only weakly attractive or even vanishing.  We believe that 
 this result remains true if $v$ is not purely attractive, but we have 
 not been able to prove this generalization.

 This paper is organized as follows.  In Sec.~\ref{Varia}, we discuss 
 how to compute accurately the critical couplings $g_{N}$.  In 
 Sec.~\ref{Resu}, we present the results obtained for the Morse model 
 and a few other potentials: the three-body binding energy obtained at 
 $g=g_{2}$, at the edge of binding for the two-body systems; the 
 critical coupling $g_{3}$ for three-body systems; an estimate of 
 $g_{4}$, the minimal strength necessary to bind four bosons.  It is 
 expected that $g_{4}<g_{3}$, with, however, the constraints 
 $g_{4}/g_{3}\ge 3/4$ and $g_{4}/g_{3}\ge 1/2$ established in 
 Ref.~\cite{Richard94}.  The numerical estimate of 
 $g_{4}$ requires delicate variational calculations, especially when 
 the potential displays both attractive and repulsive parts.  A simple 
 extrapolation to $g_{\infty}$, i.e., the infinite boson matter, will 
 be presented in Sec.~\ref{Inf-Bos-Mat}.  Some conclusions and a list 
 of open problems are presented in Sec.~\ref{Discuss}.
\section{Variational methods}\label{Varia}
There are well-known techniques, in particular variational methods 
\cite{Varga95}, to compute with very high precision the binding energy 
$E_{N}(g)$ of a system of $N$ particles in a regime $g>g_{N}$ where 
binding is established.  It is a slightly different art, however, to 
estimate the value $g_{N}$ corresponding to the border of the 
stability domain.  Even in the simple case of $N=2$ constituents, this 
is not completely obvious, as seen, e.g., from the discussion in 
Refs.~\cite{Blatt49,Gomes94} for the Yukawa potential.

A first strategy consists in computing accurately the binding 
$E_{N}(g)$ in a domain where binding occurs and letting $g$ decrease. As 
a behavior 
\begin{equation}
	E_{N}\propto -(g-g_{N})^{2}
	\label{E-behavior}
\end{equation}
is expected, one better looks at $(-E_{N})^{1/2}$ as a function of $g$ 
and checks a straight behavior as $E_{N}\to 0$.  As in Ref.  
\cite{Tanaka97}, a Pad\'e-type of approximation is found adequate to 
extrapolate towards $g_{N}$.  In a typical variational method, the 
Schr\"odinger equation $(T+V)\Psi=E_{N}\Psi$ is solved by expanding 
the wave function on a basis of functions 
\begin{equation}
	\Psi=\sum_{i}C_{i}\varphi_{i} 
	\label{Psi-expansion}
\end{equation}
In a given set of $\varphi_{i}$, the weights $C_{i}$ (represented by a vector 
$\vec{C}$) and the variational energy $E$ are obtained from a generalized 
eigenvalue equation
\begin{equation}
	(\widetilde{T}+g\widetilde{V})\vec{C}=E \widetilde{N}\vec{C},
	\label{eigenvalue-eq}
\end{equation}
involving the restrictions of the kinetic energy operator $T$ and 
potential energy $V$ to the space spanned by the $\varphi_{i}$, and a 
definite-positive matrix $\widetilde{N}$, which does not reduce to 
the unit matrix when the $\varphi_{i}$ are not orthogonal.

An alternative 
(though not strictly legal) method for estimating the critical 
coupling $g_{N}$ consists of looking directly at the point $E=0$  and 
rewrite the eigenvalue equation as
\begin{equation}
\label{Direct-g}
\widetilde{V}\vec{C}=-{1\over g}\,\widetilde{T}\vec{C},
\end{equation}
involving, again, an Hermitian matrix on the left, and a 
definite-positive matrix on the right. In principle the wave-function 
needs not be normalizable at $E=0$, but in practice, one can use a basis 
of normalizable functions, provided one allows for
components with very long range.

The results presented below have been checked using both  the 
extrapolation method  and the direct estimate of the critical coupling.

When the number of terms in Eq.~(\ref{Psi-expansion}) is incremented, 
there is a dramatic increase of the number of non-linear parameters 
entering the basis functions (the coefficients $a_{ij}$ in the 
examples below).  The minimization of the variational energy by 
varying these parameters becomes {\it i)} ambiguous, as neighboring 
sets of values give comparable energies, and {\it ii)} intractable, 
even with sophisticated minimization routines.  A simple trick 
\cite{Krikeb98} is inspired by the work of Kamimura \cite{Kamimura88}.  
It consists of imposing all $a_{ij}$ parameters to be chosen in a single 
geometric series.  Then only the lowest and the largest values have to 
be optimized numerically.  The minimization is much faster.  The 
slight loss in accuracy is more than compensated by the possibility of 
increasing easily the number of terms.  This works rather well for 
achieving a reasonable accuracy.  When one aims at very precise 
results, more sophisticated techniques are required, such as the 
well-documented and powerful stochastic variational method (SVM) 
\cite{Varga95,Suzuki98}.

It remains to choose the basis functions in Eq.~(\ref{Psi-expansion}).
We have compared the results obtained with exponential functions
\begin{equation}
	\varphi_{i}=\exp\left(-\sum a_{ij}r_{ij}\right) +\cdots 
	\label{Exponentials}
\end{equation}
and Gaussians
\begin{equation}
	\varphi_{i}=\exp\left(-\sum a_{ij}r_{ij}^{2}\right) + \cdots,
	\label{Gaussians}
\end{equation}
where the parenthesis can be rewritten as the most general quadratic 
form involving relative Jacobi coordinates.  In both cases, the dots 
are meant for terms deduced by permutation, to ensure the 
proper symmetry properties of the trial basis.

The former basis is by far more efficient when the expansion 
(\ref{Psi-expansion}) is limited to a small number of terms.  For 
instance, a single exponential function is sufficient to demonstrate 
the stability of the ion Ps$^{-}=(e^{+}e^-e^-)$ in quantum chemistry, 
while several Gaussians are needed.  However, when the number of terms 
increases, the exponential basis, even when associated with a 
stochastic search, tends to give rise to numerical instabilities 
similar to those described, e.g., by Spruch and Delves 
\cite{SpruchDelves68}.  The problem can certainly be circumvented 
\cite{Frolov}, but we found it more convenient to use SVM with 
Gaussians to get stable and accurate results.  Anyhow, the results 
involving more than $N=3$ particles have been obtained with Gaussians 
only, since one cannot derive simple analytic expressions for the 
matrix elements within the exponential basis.  Note that when $N$ 
increases, the surface and tail of the system play a relatively less 
important role, so the use of Gaussian functions should become more 
appropriate.
\section{Results for \boldmath $\lowercase{g}_{3}/\lowercase{g}_{2}$
and $\lowercase{g}_{4}/\lowercase{g}_{2}$\unboldmath}
\label{Resu}
In this section, we present some results on $R=g_{3}/g_{2}$ and 
$R_{4}=g_{4}/g_{2}$.  We restrict ourselves to symmetric 3- or 4-body 
systems, involving identical bosons.  Some results on equal-mass 
particles with asymmetric interaction have been given in 
Refs.~\cite{Richard94,Goy95}.
\subsection{Monotonous potentials}
\label{Monotonic}
We consider here three simple functional forms, Yukawa, exponential and 
Gaussian, corresponding to 
\begin{equation}
\label{YEG}
-v(r)={\exp(-\mu r)\over r},\qquad \exp(-\mu r),\qquad \exp(- \mu^{2} r^{2}),
\end{equation} 
respectively, where, without loss of generality, the range parameter 
$\mu$ and the constituent mass $m$ can be set to $\mu=m=1$ by simple 
rescaling.

The critical couplings are displayed in Table~\ref{Tab1}.  As already 
stressed, the most remarkable feature is the close clustering of all 
values of $R$ near $0.80$.  This means a 20\% window for 
Borromean 3-body binding.  Similarly, all values of 
$R_{4}$ are found around 0.64.  There is no obvious 
meaning to the observation that $g_{4}/g_{3}\simeq g_{3}/g_{2}$.  
Anyhow, $g_{N+1}/g_{N}$ cannot be smaller than $N/(N+1)$ 
\cite{Richard94}, so it should tend to 1 as $N$ increases.
\subsection{Potentials with external barrier}
The potential
\begin{equation}\label{pot-ext-bar}
	v(r)=a \exp(-\mu^{2} r^{2}/2)+b\exp(-2\mu^{2} r^{2})
\end{equation}
has been used by Nielsen et al.\ \cite{Nielsen99}, to study Borromean 
binding in two dimensions.  By rescaling, one can fix $m=\mu=1$.  The 
cases $(a,b)=(-1,0)$ and $(a,b)=(1,-2)$ are shown in Fig.~\ref{Fig1} 
for illustration.  For $a$ and $b$ both negative, this potential 
reduces to a simple monotonous function, and, not surprisingly, a 
ratio $R\simeq0.80$ is obtained for the critical couplings $g_{3}$ and 
$g_{2}$.  With $a$ and $b$ of different signs, one can build a 
potential which looks like an almost pure harmonic oscillator at small 
values of $r$ and vanishes only at distances which are very large as 
compared to the size of the ground state wave function.  One then 
obtains $R\to 2/3$.

In this limit, we are, indeed, approaching the situation where the 
decomposition
\begin{equation}
	\label{Hall-Post}
	\widetilde{H}_{3}(m,g)=\sum_{i<j}\widetilde{H}_{2}^{(i,j)}(3m/2,g)
\end{equation}
corresponds to an exact factorization of the wave function 
\cite{Basdevant90,Richard94}, and thus 
the vanishing of $\widetilde{H}_{3}(m,g)$ implies that of
$\widetilde{H}_{2}^{(i,j)}(3m/2,g)$ i.e., $g_{3}=2g_{2}/3$ \cite{Richard94}. 
Otherwise, one simply gets from Eq.~(\ref{Hall-Post}), when saturated 
with the exact 3-body wave function, $E_{3}(m,g)=\sum \langle 
\widetilde{H}_{2}\rangle\ge 3E_{2}(3m/2,g)$, i.e., $g_{3}\ge 2g_{2}/3$. 
Here
\begin{equation}\label{def-HN}
\widetilde{H}_{N}(m,g)=\sum_{k}{\vec{p}_{k}^{2}\over 
2m}+g\sum_{k<l}v(r_{kl})-{(\sum_{k}\vec{p}_{k})^{2}\over 2 N m}
\end{equation}
is the translation-invariant Hamiltonian describing the relative motion 
of $N$ particles.

Table \ref{Tab1} gives the values for a pure Gaussian potential, 
corresponding to $(a,b)=(-1,0)$: one obtains $g_{2}=2.680$, 
$g_{3}/g_{2}=0.79$ and $g_{4}/g_{3}=0.80$. For the potential 
$(a,b)=(1,-2)$ also shown in Fig.~\ref{Fig1}, the values become:
$g_{2}=21.20$, 
$g_{3}/g_{2}=0.672$ and $g_{4}/g_{3}=0.754$. We are already very close 
to the limit where $g_{N+1}/g_{N}=N/(N+1)$.

\subsection{Morse potential}
\label{Morse}
The Morse potential reads
\begin{equation}\label{Morse-Pot}
v(r)= \exp\left[-2\mu (r-r_{0})\right]-2\exp\left[-\mu(r-r_{0})\right].
\end{equation}
Again, one can set $m=\mu=1$ by rescaling.  The shape is displayed in 
Fig.~\ref{Fig2}, for $r_{0}=1$ and $r_{0}=2$.  The 2-body problem 
with this potential can be worked out exactly \cite{Fluegge71}.  In 
particular, the critical coupling $g_{2}$ is obtained from an equation 
involving the Kummer function, which can be solved easily.  Our 
normalization is such that $g_{2}\to 1/4$ as $r_{0}$ increases.  The 
critical couplings $g_{3}$ and $g_{4}$ have been estimated 
numerically, as well as $E_{3}(g_{2})$, the 3-body ground state energy 
at the edge of binding 2-body systems, and similarly $E_{4}(g_{3})$.  The results 
are shown in Table~\ref{Tab2}.

A warning is that the calculation becomes very difficult as $r_{0}$ 
becomes larger than about 3.  Our parametrization becomes inadequate.  
The vanishing of the wave function at small interparticle distance 
$r_{ij}$ is obtained at the expense of huge cancellations in the 
expansion (\ref{Psi-expansion}).  This considerably reduces the 
accuracy.  Specific methods can be developed for interactions with 
hard core, see, e.g., \cite{Kolganova98} and references therein.  Our 
results, however, seem good enough to show unambiguously the trends of 
the $R_{N}$ ratios as the size $r_{0}$ of the core increases.

The size of the attractive pocket is measured by the interval between 
$r_{1}$, where the potential vanishes, and $r_{0}$, where it reaches 
its minimum. Within our normalization, $\delta r=r_{0}-r_{1}=\ln 2$ 
is constant. As $r_{0}$ increases, $\delta r/r_{0}\to 0$, and the 
Morse potential becomes similar to the attractive delta-shell with 
hard core described in Sec.~\ref{Intro} in the limit $d/c\to1$, and 
then a behavior $R\to1$ is expected. This is clearly observed in 
Table \ref{Tab2}. The trend is, however, rather slow. For moderate 
values of $r_{0}$, our numerical results are well reproduced by a fit
\begin{equation}
\label{R-fit}
R\simeq 1-c {\delta r \over r_{0}}\simeq 1-{0.43\over r_{0}},
\end{equation}
with $c\simeq 0.62$. For larger $r_{0}$, some departure is observed, 
probably due to the difficulties in the variational calculation.
We believe the behavior (\ref{R-fit}) is rather general.
\subsection{P{\"o}schl--Teller potential}
\label{PT}
It reads
\begin{equation}
\label{PT-pot}
v(r)={\alpha(\alpha-1)\over \sinh^{2}(\mu r)}-{ 
\alpha(\alpha+1)\over\cosh^{2}(\mu r)},
\end{equation}
with, again, $\mu=m=1$ for the range and the mass of each constituent.  
The strength factors of the repulsive and attractive terms are tuned 
to give a zero-energy 2-body state, i.e., $g_{2}=1$ \cite{Fluegge71}.  
The potential is drawn in Fig.~\ref{Fig3} for some values of $\alpha$.

In the case $\alpha=1$, we have a simple monotonous potential, and, not 
surprisingly, a value $R\simeq 0.8$ is found, as seen in Table 
\ref{Tab3}.

For $\alpha>1$, we can again define the size of the attractive well as
$\delta r=r_{0}-r_{1}$, with $v(r_{1})=0$ and $ v'(r_{0})=0$. The fit 
of $R$ using the empirical formula (\ref{R-fit}) turns out to be quite 
good. This can be checked from the values displayed in Table \ref{Tab3}.

For both the Morse and P{\"o}schl--Teller potentials, the approximate 
equality of $g_{4}/g_{3}$ and $g_{3}/g_{2}$ survives a strong 
repulsive core, unlike the case of an external barrier. There is, 
however, a slight difference in the patterns exhibited by these 
potentials. In the Morse case, the ratios $g_{4}/g_{3}$ and $g_{3}/g_{2}$ 
start departing from about 0.80 at the same value of the parameter 
$r_{0}$ for which $g_{\infty}/g_{2}$ becomes positive, i.e., binding 
the infinite boson matter requires a minimal strength.
In the P{\"o}schl--Teller case, $g_{4}/g_{3}$ and $g_{3}/g_{2}$ 
immediately increase when the parameter $\alpha$ becomes larger than 1, 
though $g_{\infty}/g_{2}$ still vanishes for a while. 
\subsection{More complicated potentials}
In Ref.~\cite{Krikeb98}  the ratio $R$ is studied for the potential
\begin{equation}\label{pot-Ali}
	v(r)=r^{2}\exp(-2\mu r) -\exp(-\mu r),
	\end{equation}
 which is shown in Fig.~\ref{Fig4}, for selected values of the range 
 parameter $\mu$.  For $\mu$ very large, it is almost attractive, and 
 one thus obtains the usual $R\simeq 0.80$.  On the other hand, for 
 very small $\mu$, we have an almost pure oscillator $v(r)\simeq 
 r^{2}-1$ in the region of interest, and one gets $R\to 2/3$.  For 
 intermediate value of $\mu$, the potential exhibits both an internal 
 pocket of attraction and an external attractive tail.

The numerical results are shown in Fig.~\ref{Fig5}. The expected 
behavior $R=2/3$ for $\mu\to 0$ and $R\simeq 0.80$ for 
$\mu\to\infty$ are verified. Near $\mu=0.70$, there is an interesting 
tunnelling between the internal and the external pockets of attraction. 
The barrier is seen as an internal core by the latter, and this 
pushes $R$ toward 1, as for a Morse potential of large radius.
     
\section{Larger systems}\label{Inf-Bos-Mat}
The case of an infinite boson matter sheds some light on our 
discussion.  For a purely attractive potential, a system containing 
many bosons is bound, however weak is the strength $g$: we will thus 
set $g_{\infty}=0$.  Now, if the potential contains a large repulsive 
part, it is conceivable that binding requires a minimal strength of 
the potential, say $g>g_{\infty}>0$ to pull the wave function in the 
attractive parts of the potentials.  A result which looks {\sl a 
posteriori} reasonable, is that $g_{\infty}$ is the value of the 
coupling for which the scattering length vanishes \cite{Bruch76}.  
Indeed, the optimal state of the infinite boson matter is a compromise 
between the large-density limit, for which the kinetic energy is too 
large, and the extreme dilution.  The latter case, dominated by 
two-body collisions at zero energy, should exhibit a tendency towards 
binding, i.e., a negative scattering length.

The scattering length can be calculated analytically or numerically 
for the potentials considered previously.  Then it is rather 
straightforward to determine the value $g_{\infty}$ of the strength  which makes 
it vanish.  Of course, for a potential whose integral is negative, 
the scattering length is already negative in the weak coupling limit, 
and remains negative until $g=g_{2}$.  This corresponds to 
$g_{\infty}=0$.

In Tables \ref{Tab1},\ref{Tab2} and \ref{Tab3}, we display the value 
$g_{\infty}$ for which the scattering length vanishes.  This is simply 
$g_{\infty}=0$ for the monotonous potentials of Table \ref{Tab1} and 
the limiting cases $r_{0}=0$ of the Morse potential and $\alpha=1$ of 
P{\"o}schl--Teller.  As $r_{0}$ or $\alpha$ increases, one observes 
almost simultaneously $g_{\infty}$ becoming finite and $R=g_{3}/g_{2}$ 
departing from about 0.80 and approaching $R=1$.  This is the limit 
between, say, simple potentials which are purely attractive or contain 
a small repulsion, and non-trivial potentials with a strong core.

\section{Discussion and outlook}\label{Discuss}
In this paper, we have studied some aspects of the phenomenon of 
Borromean binding in three dimensions by comparing the critical 
couplings $g_{N}$ required for binding $N=2$, 3 or more bosons 
interacting with various types of potentials.

All monotonous, short-range potentials give almost the same ratios 
$g_{3}/g_{2}\simeq 0.80$ and $g_{4}/g_{2}\simeq 0.64$.

We then considered  potentials with a short-range attraction 
and an external repulsive barrier, which behave  very much like an 
oscillator, and, not surprisingly, give ratios of critical couplings 
close to the lower bound $g_{N+1}/g_{N}=N/(N+1)$. These potentials 
with an external barrier are not very often encountered in physical 
systems. They are, however, interesting, since, according to 
Ref.~\cite{Nielsen99}, they are the only ones to give rise to 
Borromean binding in two dimensions.

We then studied the more physical case of potentials with a strong 
repulsive core at short distances.  The window for Borromean binding 
turns out to be much narrower than for purely attractive potentials.

The present investigation could be extended to excited states. In 
particular, as long as $g_{3}<g_{2}$, the Efimov effect should remain as $g$ 
approaches $g_{2}$. It would be interesting to study how the onset and 
disappearance of Efimov states change when the strength of the core 
is varied.

\acknowledgments This investigation was stimulated by discussions 
during several visits of two of us (S.M. and J.M.R.) at ECT$^{*}$ 
(European Centre for Theoretical Nuclear Physics and Related Areas).  
We would like to thank ECT$^*$ for the generous hospitality provided 
to us.  
%
%
%

\listoftables
\begin{table}
\caption{\label{Tab1} Comparison of critical couplings $g_{N}$ to 
achieve binding of $N=2$, 3 and 4 identical bosons in a Yukawa (Y), 
exponential (E) or Gaussian (G) potential.  Also shown is $g_{\infty}$ 
which correspond to a vanishing of the 2-body scattering length, and 
the 3-body energy $E_{3}$ obtained for $g=g_{2}$, and the 4-body 
energy $E_{4}$ for $g=g_{3}$.}
\begin{tabular}{cdddddd}
Potential&
$g_{2}$&
$E_{3}(g_{2})$&
$g_{3}/g_{2}$&
$E_{4}(g_{3})$&
$g_{4}/g_{3}$&    
$g_{\infty}/g_{2}$\\
\hline
Y& 1.68 &--0.172   &0.80  &--0.320     &0.81   
                                     &0      \\
E& 1.45 &--0.047   &0.80  &--0.093     &0.80   
                                     &0      \\
G& 2.68 &--0.236   &0.79  &--0.438    &0.80   
                                     &0      \\
\end{tabular}
\end{table}
\begin{table}
\caption{\label{Tab2} Same as Table \protect\ref{Tab1}, but for a 
Morse potential (\protect\ref{Morse-Pot}), whose characteristic radius 
$r_{0}$ is varied.}
\begin{tabular}{ddddddd}
$r_{0}$&
$g_{2}$&
$E_{3}(g_{2})$&
$g_{3}/g_{2}$&
$E_{4}(g_{3})$&
$g_{4}/g_{3}$&    
$g_{\infty}/g_{2}$\\
\hline
0.  &0.810 &--0.0411  & 0.799&--0.0808   &0.798      
                                              &0\\
1.0 &0.369 &--0.0325  & 0.797&--0.0636   &0.790      
                                              &0\\
2.0 &0.254 &--0.0174  & 0.807&--0.0333   &0.794      
                                              &0\\
3.0 &0.250 &--0.0081  & 0.862&--0.0146   &0.860      
                                              &0.09\\
4.0 &0.250 &--0.0046  & 0.900&--0.0080   &0.907      
                                              &0.28\\
\end{tabular}
\end{table}
\begin{table}
\caption{\label{Tab3} Same as Table \protect\ref{Tab1}, but for a 
P{\"o}schl--Teller potential (\protect\ref{PT-pot}), for several 
values of its parameter $\alpha$.}
\begin{tabular}{ddddddd}
$\alpha$&
$g_{2}$&
$E_{3}(g_{2})$&
$g_{3}/g_{2}$&
$E_{4}(g_{3})$&
$g_{4}/g_{3}$&    
$g_{\infty}/g_{2}$ \\
\hline
1&1.&--0.135 & 0.797&--0.264   &0.796          
                                              &0\\
2&1.&--0.064 & 0.818&--0.131   &0.777          
                                              &0\\
3&1.&--0.046 & 0.836&--0.085   &0.835          
                                              &0\\
5&1.&--0.032 & 0.859&--0.060   &0.856          
                                              &0.12\\
9&1.&--0.018 & 0.885&--0.042   &0.878          
                                              &0.23\\
\end{tabular}
\end{table}
\listoffigures
    \begin{figure}[htbc]
  \vskip 1cm
 \begin{center}
\setlength{\unitlength}{40pt}
  \begin{picture}(4.5,3.5)(-1,-2.5)
  \font\gnuplot=cmr10 at 12pt
  \gnuplot
  \linethickness{.7pt}
  \put(0,0){\vector(1,0){4}}
  \put(0,-2.2){\vector(0,1){3}}
\put(-.09,0){\makebox(0,0)[r]{0}}
\put(0,-2){\makebox(0,0)[c]{$\scriptscriptstyle -$}}
\put(-.09,-2){\makebox(0,0)[r]{-1}}
\put(1,0){\makebox(0,0)[c]{$\scriptscriptstyle \vert$}}
\put(2,0){\makebox(0,0)[c]{$\scriptscriptstyle \vert$}}
\put(3,0){\makebox(0,0)[c]{$\scriptscriptstyle \vert$}}
\put(1,.20){\makebox(0,0)[c]{1}}
\put(3,-.20){\makebox(0,0)[c]{3}}
\put(0,0.9){\makebox(0,0)[r]{$V(r)$}}
\put(4.1,0){\makebox(0,0)[l]{$r$}}
\linethickness{1.0pt}
\qbezier[50](0,-2.)(0.176,-2.)(0.364,-1.87)
\qbezier[50](0.364,-1.87)(0.522,-1.76)(0.727,-1.54)
\qbezier[50](0.727,-1.54)(0.79,-1.47)(1.09,-1.1)
\qbezier[50](1.09,-1.1)(1.31,-0.845)(1.45,-0.694)
\qbezier[50](1.45,-0.694)(1.64,-0.507)(1.82,-0.383)
\qbezier[50](1.82,-0.383)(1.99,-0.261)(2.18,-0.185)
\qbezier[50](2.18,-0.185)(2.35,-0.118)(2.55,-0.0784)
\qbezier[50](2.55,-0.0784)(2.71,-0.0462)(2.91,-0.0291)
\qbezier[50](2.91,-0.0291)(3.07,-0.0159)(3.27,-0.00945)
\qbezier[50](3.27,-0.00945)(3.42,-0.00476)(3.64,-0.00269)
\qbezier[50](3.64,-0.00269)(3.78,-0.00125)(4.,-0.000671)
\qbezier[50](0,-2.)(0.0999,-2.)(0.211,-1.7)
\qbezier[50](0.211,-1.7)(0.291,-1.49)(0.421,-0.976)
\qbezier[50](0.421,-0.976)(0.586,-0.323)(0.632,-0.163)
\qbezier[50](0.632,-0.163)(0.743,0.227)(0.842,0.434)
\qbezier[50](0.842,0.434)(0.943,0.644)(1.05,0.713)
\qbezier[50](1.05,0.713)(1.15,0.772)(1.26,0.736)
\qbezier[50](1.26,0.736)(1.35,0.71)(1.47,0.623)
\qbezier[50](1.47,0.623)(1.46,0.634)(1.68,0.471)
\qbezier[50](1.68,0.471)(1.8,0.386)(1.89,0.329)
\qbezier[50](1.89,0.329)(2.,0.265)(2.11,0.218)
\qbezier[50](2.11,0.218)(2.21,0.171)(2.32,0.137)
\qbezier[50](2.32,0.137)(2.42,0.105)(2.53,0.0822)
\qbezier[50](2.53,0.0822)(2.63,0.0617)(2.74,0.0473)
\qbezier[50](2.74,0.0473)(2.83,0.0346)(2.95,0.026)
\qbezier[50](2.95,0.026)(3.04,0.0186)(3.16,0.0137)
\qbezier[50](3.16,0.0137)(3.25,0.00953)(3.37,0.00687)
\qbezier[50](3.37,0.00687)(3.46,0.00468)(3.58,0.00331)
\qbezier[50](3.58,0.00331)(3.67,0.0022)(3.79,0.00152)
\qbezier[50](3.79,0.00152)(3.88,0.000986)(4.,0.000671)
\end{picture}
\caption{Shape of the potential of Eq.~(\protect\ref{pot-ext-bar}) for $(a,b)=(1,-2)$ (with 
external barrier) and $(-1,0)$ (monotonous).
\label{Fig1}}  
\end{center}
\end{figure}
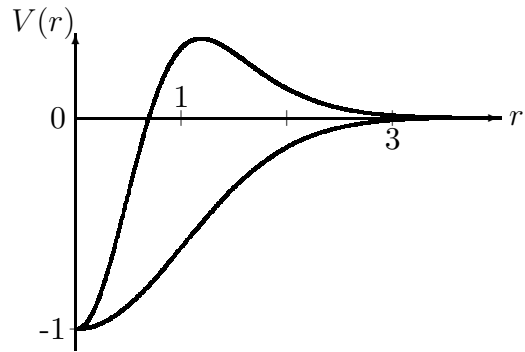    
\begin{figure}[htbc]
  \vskip 1cm
 \begin{center}
 \setlength{\unitlength}{40pt}
  \begin{picture}(4.5,3.5)(-1,-2)
  \font\gnuplot=cmr10 at 12pt
  \gnuplot
  \linethickness{.7pt}
  \put(0,0){\vector(1,0){4}}
  \put(0,-1){\vector(0,1){3}}
\put(-.09,0){\makebox(0,0)[r]{0}}
\put(0,1){\makebox(0,0)[c]{$\scriptscriptstyle -$}}
\put(-.09,1){\makebox(0,0)[r]{1}}
\put(0,-1){\makebox(0,0)[c]{$\scriptscriptstyle -$}}
\put(-.09,-1){\makebox(0,0)[r]{-1}}
\put(1,0){\makebox(0,0)[c]{$\scriptscriptstyle \vert$}}
\put(1,.2){\makebox(0,0)[c]{1}}
\put(2,0){\makebox(0,0)[c]{$\scriptscriptstyle \vert$}}
\put(2,.2){\makebox(0,0)[c]{2}}
\put(3,0){\makebox(0,0)[c]{$\scriptscriptstyle \vert$}}
\put(3,.2){\makebox(0,0)[c]{3}}
\put(0,2.1){\makebox(0,0)[r]{$V(r)$}}
\put(4.1,0){\makebox(0,0)[l]{$r$}}
\linethickness{1.0pt}
\qbezier[150](0.001,1.94)(0.147,0.585)(0.334,-0.105)
\qbezier[150](0.334,-0.105)(0.479,-0.637)(0.668,-0.845)
\qbezier[50](0.668,-0.845)(0.809,-1.)(1.,-1.)
\qbezier[50](1.,-1.)(1.14,-1.)(1.33,-0.919)
\qbezier[50](1.33,-0.919)(1.44,-0.875)(1.67,-0.763)
\qbezier[50](1.67,-0.763)(1.89,-0.65)(2.,-0.6)
\qbezier[50](2.,-0.6)(2.17,-0.52)(2.33,-0.457)
\qbezier[50](2.33,-0.457)(2.5,-0.393)(2.67,-0.342)
\qbezier[50](2.67,-0.342)(2.83,-0.292)(3.,-0.252)
\qbezier[50](3.,-0.252)(3.16,-0.215)(3.33,-0.184)
\qbezier[50](3.33,-0.184)(3.49,-0.157)(3.67,-0.134)
\qbezier[150](0.995,2.)(1.11,0.926)(1.25,0.269)
\qbezier[150](1.25,0.269)(1.36,-0.272)(1.5,-0.571)
\qbezier[50](1.5,-0.571)(1.61,-0.814)(1.75,-0.917)
\qbezier[50](1.75,-0.917)(1.86,-0.999)(2.,-1.)
\qbezier[50](2.,-1.)(2.11,-1.)(2.25,-0.952)
\qbezier[50](2.25,-0.952)(2.35,-0.917)(2.5,-0.846)
\qbezier[50](2.5,-0.846)(2.55,-0.822)(2.75,-0.722)
\qbezier[50](2.75,-0.722)(2.9,-0.649)(3.,-0.601)
\qbezier[50](3.,-0.601)(3.13,-0.54)(3.25,-0.491)
\qbezier[50](3.25,-0.491)(3.37,-0.44)(3.5,-0.396)
\qbezier[50](3.5,-0.396)(3.62,-0.354)(3.75,-0.317)
\end{picture}
\caption{Shape of the Morse potential for $r_{0}=1$ and $r_{0}=2$.
\label{Fig2}}  
\end{center}
\end{figure}
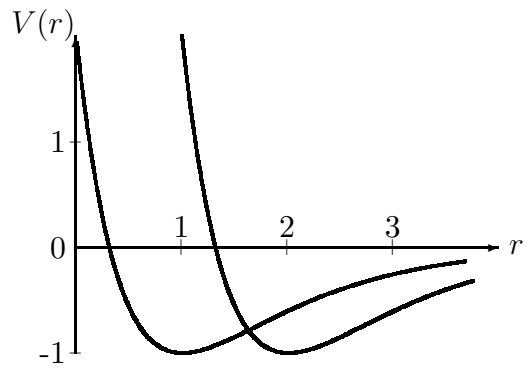
\begin{figure}[htbc]
  \vskip 1cm
 \begin{center}
 \setlength{\unitlength}{40pt}
  \begin{picture}(4.5,3.9)(-1,-2)
  \font\gnuplot=cmr10 at 12pt
  \gnuplot
  \linethickness{.7pt}
  \put(0,0){\vector(1,0){4}}
  \put(0,-1){\vector(0,1){3}}
\put(-.09,0){\makebox(0,0)[r]{0}}
\put(0,1){\makebox(0,0)[c]{$\scriptscriptstyle -$}}
\put(-.09,1){\makebox(0,0)[r]{2}}
\put(0,-1){\makebox(0,0)[c]{$\scriptscriptstyle -$}}
\put(-.09,-1){\makebox(0,0)[r]{-2}}
\put(1,0){\makebox(0,0)[c]{$\scriptscriptstyle \vert$}}
\put(1,.2){\makebox(0,0)[c]{1}}
\put(2,0){\makebox(0,0)[c]{$\scriptscriptstyle \vert$}}
\put(2,.2){\makebox(0,0)[c]{2}}
\put(3,0){\makebox(0,0)[c]{$\scriptscriptstyle \vert$}}
\put(3,.2){\makebox(0,0)[c]{3}}
\put(0,2.1){\makebox(0,0)[r]{$V(r)$}}
\put(4.1,0){\makebox(0,0)[l]{$r$}}
\linethickness{1.0pt}
\qbezier[50](0.001,-1.)(0.155,-1.)(0.333,-0.897)
\qbezier[50](0.333,-0.897)(0.439,-0.836)(0.666,-0.661)
\qbezier[50](0.666,-0.661)(0.874,-0.5)(0.998,-0.421)
\qbezier[50](0.998,-0.421)(1.16,-0.315)(1.33,-0.244)
\qbezier[50](1.33,-0.244)(1.49,-0.178)(1.66,-0.134)
\qbezier[50](1.66,-0.134)(1.81,-0.0961)(2.,-0.0713)
\qbezier[50](2.,-0.0713)(2.15,-0.0507)(2.33,-0.0373)
\qbezier[50](2.33,-0.0373)(2.48,-0.0264)(2.66,-0.0194)
\qbezier[50](2.66,-0.0194)(2.81,-0.0137)(2.99,-0.01)
\qbezier[50](2.99,-0.01)(3.14,-0.00706)(3.33,-0.00516)
\qbezier[50](3.33,-0.00516)(3.47,-0.00364)(3.66,-0.00266)
\qbezier[50](3.66,-0.00266)(3.81,-0.00187)(3.99,-0.00137)
\qbezier[250](0.458,2.)(0.558,0.139)(0.752,-0.316)
\qbezier[50](0.752,-0.316)(0.859,-0.566)(1.05,-0.531)
\qbezier[50](1.05,-0.531)(1.13,-0.515)(1.34,-0.404)
\qbezier[50](1.34,-0.404)(1.52,-0.307)(1.64,-0.259)
\qbezier[50](1.64,-0.259)(1.78,-0.197)(1.93,-0.155)
\qbezier[50](1.93,-0.155)(2.07,-0.116)(2.22,-0.0893)
\qbezier[50](2.22,-0.0893)(2.36,-0.0663)(2.52,-0.0506)
\qbezier[50](2.52,-0.0506)(2.65,-0.0374)(2.81,-0.0284)
\qbezier[50](2.81,-0.0284)(2.95,-0.021)(3.11,-0.0159)
\qbezier[50](3.11,-0.0159)(3.24,-0.0117)(3.4,-0.00885)
\qbezier[50](3.4,-0.00885)(3.53,-0.0065)(3.7,-0.00492)
\qbezier[50](3.7,-0.00492)(3.83,-0.00362)(3.99,-0.00273)
\qbezier[250](0.659,1.99)(0.763,0.315)(0.937,-0.197)
\qbezier[50](0.937,-0.197)(1.04,-0.507)(1.21,-0.515)
\qbezier[50](1.21,-0.515)(1.31,-0.518)(1.49,-0.427)
\qbezier[50](1.49,-0.427)(1.77,-0.29)(1.77,-0.289)
\qbezier[50](1.77,-0.289)(1.91,-0.224)(2.05,-0.18)
\qbezier[50](2.05,-0.18)(2.18,-0.138)(2.32,-0.108)
\qbezier[50](2.32,-0.108)(2.45,-0.0821)(2.6,-0.0638)
\qbezier[50](2.6,-0.0638)(2.73,-0.0481)(2.88,-0.0371)
\qbezier[50](2.88,-0.0371)(3.01,-0.0279)(3.16,-0.0215)
\qbezier[50](3.16,-0.0215)(3.28,-0.0161)(3.43,-0.0124)
\qbezier[50](3.43,-0.0124)(3.56,-0.00928)(3.71,-0.00713)
\qbezier[50](3.71,-0.00713)(3.84,-0.00534)(3.99,-0.0041)
\qbezier[250](0.912,2.)(1.01,0.464)(1.17,-0.085)
\qbezier[50](1.17,-0.085)(1.27,-0.444)(1.42,-0.495)
\qbezier[50](1.42,-0.495)(1.52,-0.526)(1.68,-0.455)
\qbezier[50](1.68,-0.455)(1.68,-0.458)(1.94,-0.329)
\qbezier[50](1.94,-0.329)(2.08,-0.262)(2.19,-0.218)
\qbezier[50](2.19,-0.218)(2.32,-0.171)(2.45,-0.138)
\qbezier[50](2.45,-0.138)(2.57,-0.107)(2.71,-0.085)
\qbezier[50](2.71,-0.085)(2.83,-0.0657)(2.96,-0.0519)
\qbezier[50](2.96,-0.0519)(3.08,-0.0399)(3.22,-0.0314)
\qbezier[50](3.22,-0.0314)(3.34,-0.0241)(3.48,-0.0189)
\qbezier[50](3.48,-0.0189)(3.59,-0.0145)(3.73,-0.0114)
\qbezier[50](3.73,-0.0114)(3.85,-0.00871)(3.99,-0.00682)
\qbezier[250](1.21,1.99)(1.3,0.591)(1.44,0.0231)
\qbezier[50](1.44,0.0231)(1.53,-0.37)(1.67,-0.465)
\qbezier[50](1.67,-0.465)(1.76,-0.527)(1.9,-0.481)
\qbezier[50](1.9,-0.481)(1.97,-0.459)(2.13,-0.377)
\qbezier[50](2.13,-0.377)(2.28,-0.304)(2.37,-0.267)
\qbezier[50](2.37,-0.267)(2.48,-0.216)(2.6,-0.18)
\qbezier[50](2.6,-0.18)(2.71,-0.144)(2.83,-0.118)
\qbezier[50](2.83,-0.118)(2.94,-0.0935)(3.06,-0.0757)
\qbezier[50](3.06,-0.0757)(3.17,-0.06)(3.29,-0.0483)
\qbezier[50](3.29,-0.0483)(3.4,-0.0382)(3.53,-0.0307)
\qbezier[50](3.53,-0.0307)(3.63,-0.0242)(3.76,-0.0194)
\qbezier[50](3.76,-0.0194)(3.87,-0.0153)(3.99,-0.0122)
\end{picture} 
\caption{Shape of the P{\"o}schl--Teller potential for 
$\alpha=1$ (monotonous) and, from left to right $\alpha=2,\,3,\,5$ and $9$.
\label{Fig3}}
\end{center}
\end{figure}
\begin{figure}[htbc]
  \vskip 1cm
 \begin{center}
\setlength{\unitlength}{30pt}
  \begin{picture}(7.5,5.5)(-1,-3)
  \font\gnuplot=cmr10 at 12pt
  \gnuplot
\linethickness{.7pt}
  \put(0,0){\vector(1,0){6.2}}
  \put(0,-2.5){\vector(0,1){4.8}}
  \put(6.5,0){\makebox(0,0)[r]{$r$}}
  \put(-0.0,2.6){\makebox(0,0)[c]{$V$}}
 \put(1,0){\makebox(0,0)[c]{$\scriptscriptstyle \vert$}}
 \put(2,0){\makebox(0,0)[c]{$\scriptscriptstyle \vert$}}
 \put(3,0){\makebox(0,0)[c]{$\scriptscriptstyle \vert$}}
 \put(4,0){\makebox(0,0)[c]{$\scriptscriptstyle \vert$}}
 \put(5,0){\makebox(0,0)[c]{$\scriptscriptstyle \vert$}} 
 \put(6,0){\makebox(0,0)[c]{$\scriptscriptstyle \vert$}} 
 \put(-.20,0){\makebox(0,0)[r]{0}}
 \put(1,.3){\makebox(0,0)[c]{1}}
  \put(4,-.3){\makebox(0,0)[c]{4}}
    \put(5,-.3){\makebox(0,0)[c]{5}}
      \put(6,-.3){\makebox(0,0)[c]{6}}
 \put(0,-2){\makebox(0,0)[c]{$\scriptscriptstyle -$}}
 \put(-.2,-2){\makebox(0,0)[r]{$-1$}}
 \put(0,2){\makebox(0,0)[c]{$\scriptscriptstyle -$}}
 \put(-.2,2){\makebox(0,0)[r]{1}}
 \linethickness{1.2pt}
\qbezier[50](0,-2.)(0.186,-1.85)(0.545,-1.22)
\qbezier[50](0.545,-1.22)(0.956,-0.507)(1.09,-0.298)
\qbezier[50](1.09,-0.298)(1.37,0.135)(1.64,0.407)
\qbezier[50](1.64,0.407)(1.9,0.677)(2.18,0.826)
\qbezier[50](2.18,0.826)(2.44,0.961)(2.73,1.01)
\qbezier[50](2.73,1.01)(2.98,1.05)(3.27,1.02)
\qbezier[50](3.27,1.02)(3.51,1.)(3.82,0.94)
\qbezier[50](3.82,0.94)(4.04,0.895)(4.36,0.811)
\qbezier[50](4.36,0.811)(4.19,0.856)(4.91,0.669)
\qbezier[50](4.91,0.669)(5.22,0.588)(5.45,0.532)
\qbezier[50](5.45,0.532)(5.74,0.465)(6.,0.411)
 \linethickness{0.5pt}
\qbezier[10](0,-2.)(-0.109,-2.15)(0.545,-1.1)
\qbezier[10](0.545,-1.1)(0.825,-0.655)(1.09,-0.411)
\qbezier[10](1.09,-0.411)(1.34,-0.182)(1.64,-0.0704)
\qbezier[10](1.64,-0.0704)(1.87,0.019)(2.18,0.048)
\qbezier[10](2.18,0.048)(2.4,0.069)(2.73,0.0632)
\qbezier[10](2.73,0.0632)(2.91,0.0599)(3.27,0.0436)
\qbezier[10](3.27,0.0436)(3.67,0.0257)(3.82,0.02)
\qbezier[10](3.82,0.02)(4.1,0.0092)(4.36,0.0025)
\qbezier[10](4.36,0.0025)(4.62,-0.00398)(4.91,-0.00756)
\qbezier[10](4.91,-0.00756)(5.16,-0.0107)(5.45,-0.0119)
\qbezier[10](5.45,-0.0119)(5.7,-0.0129)(6.,-0.0127)
\linethickness{.5pt}
\qbezier[50](0,-2.)(0.447,-1.11)(0.545,-0.959)
\qbezier[50](0.545,-0.959)(0.793,-0.59)(1.09,-0.403)
\qbezier[50](1.09,-0.403)(1.32,-0.26)(1.64,-0.186)
\qbezier[50](1.64,-0.186)(1.86,-0.135)(2.18,-0.104)
\qbezier[50](2.18,-0.104)(2.41,-0.0833)(2.73,-0.0672)
\qbezier[50](2.73,-0.0672)(2.97,-0.0551)(3.27,-0.045)
\qbezier[50](3.27,-0.045)(3.53,-0.0366)(3.82,-0.0299)
\qbezier[50](3.82,-0.0299)(4.08,-0.0238)(4.36,-0.0193)
\qbezier[50](4.36,-0.0193)(4.62,-0.0152)(4.91,-0.0121)
\qbezier[50](4.91,-0.0121)(5.16,-0.00943)(5.45,-0.00747)
\qbezier[50](5.45,-0.00747)(5.71,-0.00575)(6.,-0.00452)
\end{picture}
\end{center}
\caption{\label{Fig4} Potential of Eq.~(\protect\ref{pot-Ali}), for 
$\mu=0.4$ (thick line), $\mu=1$ (thin line) and $\mu=0.65$ (dotted 
line).  The potential is always negative at large distance $r$.}
\end{figure}
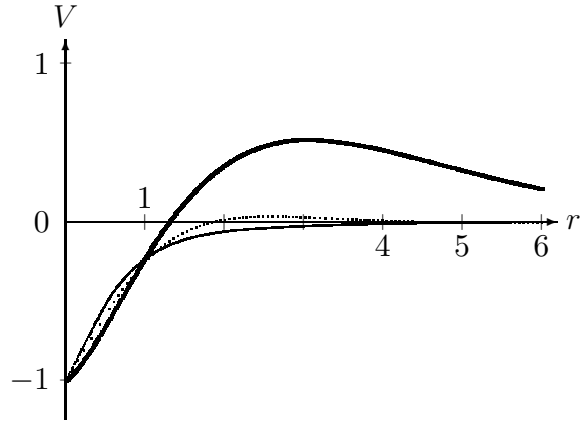
    
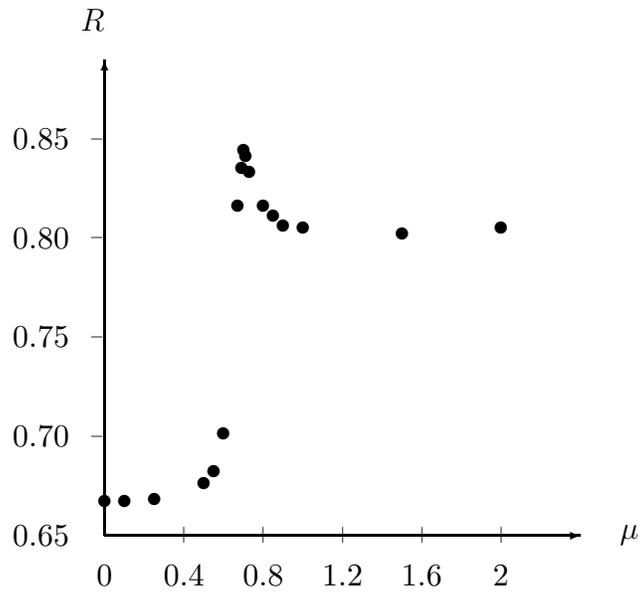
\begin{figure}[htbc]
  \vskip 1cm
 \begin{center}
\setlength{\unitlength}{150pt}
  \begin{picture}(1.5,1.5)(-.2,-.2)
  \font\gnuplot=cmr10 at 12pt
  \gnuplot
\linethickness{.7pt}
  \put(0,0){\vector(1,0){1.2}}
  \put(0,0){\vector(0,1){1.2}}
  \put(0,1.3){\makebox(0,0)[r]{$R$}}
  \put(1.3,0){\makebox(0,0)[l]{$\mu$}}
\put(.2,0){\makebox(0,0)[c]{$\scriptscriptstyle \vert$}}
\put(.4,0){\makebox(0,0)[c]{$\scriptscriptstyle \vert$}}
\put(.6,0){\makebox(0,0)[c]{$\scriptscriptstyle \vert$}}
\put(.8,0){\makebox(0,0)[c]{$\scriptscriptstyle \vert$}}
\put(1,0){\makebox(0,0)[c]{$\scriptscriptstyle \vert$}} 
\put(0,-.1){\makebox(0,0)[c]{0}}
\put(0.2,-.1){\makebox(0,0)[c]{0.4}}
\put(0.4,-.1){\makebox(0,0)[c]{0.8}}
\put(0.6,-.1){\makebox(0,0)[c]{1.2}}
\put(0.8,-.1){\makebox(0,0)[c]{1.6}}
\put(1,-.1){\makebox(0,0)[c]{2}}
\put(-.09,0){\makebox(0,0)[r]{0.65}}
\put(0,.25){\makebox(0,0)[r]{$\scriptscriptstyle -$}}
\put(-.09,0.25){\makebox(0,0)[r]{0.70}}
\put(0,.50){\makebox(0,0)[r]{$\scriptscriptstyle -$}}
\put(-.09,0.50){\makebox(0,0)[r]{0.75}}
\put(0,.75){\makebox(0,0)[r]{$\scriptscriptstyle -$}}
\put(-.09,0.75){\makebox(0,0)[r]{0.80}}
\put(0,1){\makebox(0,0)[r]{$\scriptscriptstyle -$}}
\put(-.09,1){\makebox(0,0)[r]{0.85}}
\put(0,0.085){\makebox(0,0)[c]{$\bullet$}}
\put(0.05,0.085){\makebox(0,0)[c]{$\bullet$}}
\put(0.125,0.09){\makebox(0,0)[c]{$\bullet$}}
\put(0.25,0.13){\makebox(0,0)[c]{$\bullet$}}
\put(0.275,0.16){\makebox(0,0)[c]{$\bullet$}}
\put(0.30,0.255){\makebox(0,0)[c]{$\bullet$}}
\put(0.335,0.83){\makebox(0,0)[c]{$\bullet$}}
\put(0.345,0.925){\makebox(0,0)[c]{$\bullet$}}
\put(0.35,0.97){\makebox(0,0)[c]{$\bullet$}}
\put(0.355,0.955){\makebox(0,0)[c]{$\bullet$}}
\put(0.365,0.915){\makebox(0,0)[c]{$\bullet$}}
\put(0.40,0.83){\makebox(0,0)[c]{$\bullet$}}
\put(0.425,0.805){\makebox(0,0)[c]{$\bullet$}}
\put(0.45,0.78){\makebox(0,0)[c]{$\bullet$}}
\put(0.5,0.775){\makebox(0,0)[c]{$\bullet$}}
\put(0.75,0.76){\makebox(0,0)[c]{$\bullet$}}
\put(1,0.775){\makebox(0,0)[c]{$\bullet$}}
\end{picture}
\end{center}
\caption{\label{Fig5} Computed value of $R=g_{3}/g_{2}$ for the 
potential of Eq.~(\protect\ref{pot-Ali}), for various values of the range 
parameter $\mu$.}
\end{figure}
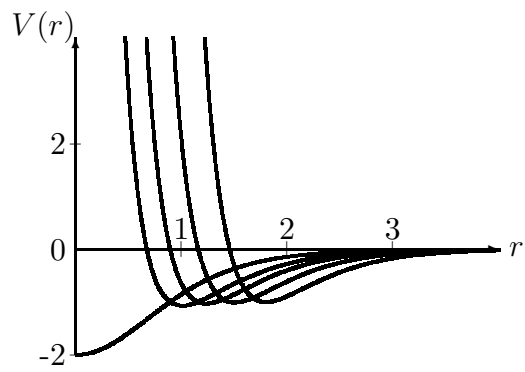
\end{document}